# Models of the Mass-Ejection Histories of pre Planetary Nebulae, III. The Shaping of Lobes by post-AGB Winds

Bruce Balick[1], Adam Frank[2] and Baowei Liu[3]

[1] Department of Astronomy, University of Washington, Seattle, WA 98195-1580, USA
[2] Departments of Physics and Astronomy, University of Rochester, Rochester, NY 14627, USA
[3] Center for Integrated Research Computing, University of Rochester, Rochester, NY 14627, USA
E-mail: balick@uw.edu

**Abstract**

We develop a physical framework for interpreting high-resolution images of pre planetary nebule ("prePNe") with pairs of candle shaped lobes. We use hydrodynamical models to infer the historical properties of the flows injected from the nucleus that shape the lobes into standard forms. First, we find a suitable set of parameters of a fast, collimated, tapered flow that is actively reshaped by an exterior slow AGB wind and that nicely fits the basic shape, kinematics, mass, and momenta of this class of prePNe. Next we vary the most influential parameters of this "baseline" model-such as density, speed, and geometry- to see how changes in the flow parameters affect the nebular observables after 900y. Several generic conulsions emerge, such as the injected flows that create the hollow candle-shaped lobes must be light, "tapered", and injected considerably faster than the lobe expansion speed. Multi-polar and starfish prePNe probably evolve from wide angle flows in which thin-shell instabilites corrugate their leading edges. We show how the common linear relationship of Doppler shift and position along the lobe is a robust outcome the interaction of tapered diverging streamlines with the lobes' curved walls. Finally we probe how magnetic fields affect the basline model by adding a toroidal field to the injected baseline flow. Examples of prePNe and PNe that may have been magnetically shaped are listed. We conclude that the light, field-free, tapered baseline flow model is an successful and universal pardigm for unravelling the histories of lobe formation in prePNe.

Keywords: planetary nebulae: general, stars: AGB and post-AGB, stars: winds, outflows, ISM: jets and outflows

## 1. Introduction

Pre planetary nebulae ("prePNe") are a formed shortly after the end of the AGB evolutionary phase. The gas from which PNe eventually form is released in a collimated flow, often with a very high degree of symmetry whose origins suggest that a collimated flow is injected into the pre-existing isotropic ejecta of the former AGB star (Höfner and Olofsson 2018A&ARv..26....1H).

The collimated flow is shaped as it enters the much slower ambient environment downstream (long before the nebula is ionized by the still-emerging hot white dwarf in the core). For its first $\sim 10^3$y the neutral flow is compact and difficult to observe in detail without the most modern high-resolution imaging tools with sensitive detectors.

A gallery of the brighter and most symmetric prePNe at an age of ~1000y are shown in Fig. 1. All of them are members of a large subclass of prePNe known as "bipolars" for the pairs of lobes on their symmetry axes. Our primary science goal is to recreate the histories of the lobes' evolution to place hard constraints on models of their formation.

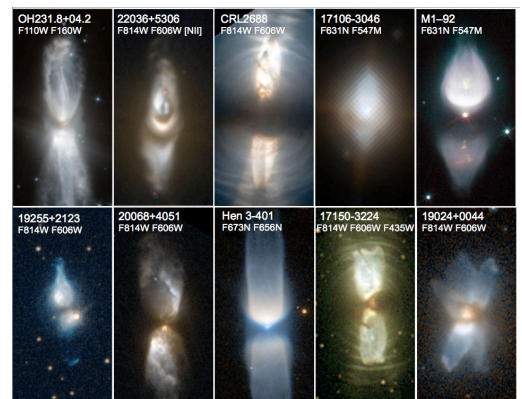

Fig 1. Images of a large subclass of prePNe with candle-shaped lobes. These images of dust-scattered starlight were reconstructed from wide-band images from the Hubble Space Telescope ("HST") in the Mikulski Archives. The target names are IRAS designations unless otherwise shown. The image of M1-92 is courtesy ESA/Hubble. A much more exhaustive catalogue of prePN images of all types can be found in Sahai et al., 2011AJ....141..134S.

In this paper we present parameter studies of hydrodynamic models to follow the generic features of lobes such as those of Fig. 1 into the past when the lobes were in their infancy. These hydrodynamic simulations, coupled with new observational tools coming on line in the next ten years, provide a unique and propitious opportunity to explore the mechanisms of collimated post-AGB mass loss. Our intent to characterize the geometry, density, and speed of the flow at a radius *r* of 1 kau (1000 au, or 2″ at a distance of 2 kpc) using realistic and tightly data constrained hydrodynamic models of the flow that match observations of the lobes in their present state.

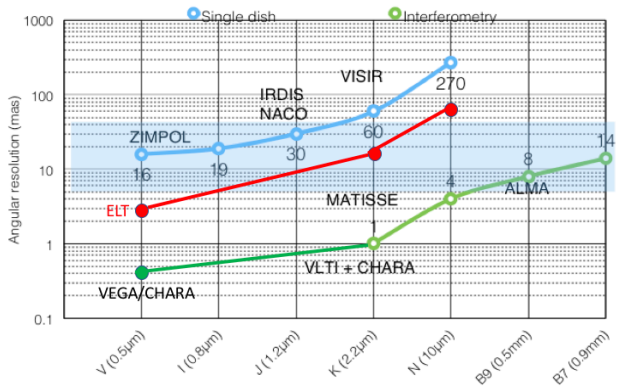

Fig. 2. New high-resolution astronomical imaging instruments anticipated over the next decade (courtesy Eric Lagadec, used by permission).

The birth processes of PNe are slowly being revealed by theoretical simulations[1] for which substabtive observational constraints are as yet unavailable. This will change as an exciting array of extemely high-spatial-resolution instruments (Fig. 2) come into operation. The new images will both enhance our models and possibly present new shaping paradigms.

The current ideas of PN formation date back to Morris, (1987PASP...99.1115M) and took their present forms by about 2000 (see the review by Balick & Frank 2002ARA&A..40..439B). Many hydrodynamic simulations of prePNe have appeared after this review was published. Based on patterns gleaned from the outflows of OH/IR stars, Zijlstra et al. (2001MNRAS.322..280Z) developed the first analytical description of bipolar lobes formed by polar flows into AGB winds. Since then various numerical simulations of bipolar prePNe have appeared. Here we mention a few with the greatest impact. Soker (2002ApJ...568..726S) and Lee & Sahai (2003ApJ...586..319L; hereafter "LS03") ran hydrodynamic models of lobes created by collimated fast winds, or "jets", that penetrate into the slower and denser AGB wind surrounding the nucleus. LS03 showed that low-density "tapered" conical steady flows, or "sprays" (section 2) injected at constant speed will encounter and shock the displaced ambient gas and develop a quasi-linear Doppler speed vs. distance relation along the lobes' walls. (Their very important result has not been appropriately recognized.)

Akashi & Soker (2008MNRAS.391. 1063A) examined short-duration jets (aka "bullets" and "clumps") using a very general numerical methodology. Their paper contains some significant insights into large-scale shapes and the onset of surface instabillities that appear along the outer boundaries of lobes. They were also the first to define the concepts of "light" and "heavy" collimated flows. A light (heavy) flow is one in which the flow density is less than (greater than) that of the local ambient medium. Light flows produce hollow lobes bordered by dense walls of displaced gas. The flows interact with and can be strongly influenced by the density distribution of their surroundings. The expanding wind-formed immediately begin to accrete mass from the downstream gas and decelerate. Moreover, the flow sreamlines do not remain radial once they striker the walls. The lobes expand homologously once they move into the dilute outer regions of the ambient gas.

Hydrodynamic models of specific nebulae are appearing frequently. For example, our group has used numerical simulations like those reported in this paper to account for the detailed structures and to retrace the evolution of three prePNe, CRL2688 (Balick et al., 2013ApJ...772...20B), OH231.8+04.2 (Balick et al., 2017ApJ...843..108B), and M29 (Balick et al., 2018ApJ...853..168B). However, the authors did not attempt to generalize from these case studies.

We shall develop a generic interpretive framework in order to probe the flow condiytions early in their evolution. In section 2 we develop a synthetic "baseline" nebula that nicely characterizes not only the images, but also the kinematics, masses, and gemoetries of prePNe. In doing so we define a set of fundamental descriptive model parameters, including the flow density, speed, opening angle, and toroidal magnetic fields, that are varied to asses how changes in these parameters influence the evolution and shapes of the baseline model. The parameter studies appear in sections 3 and 4. The observed trends of kinematic observations of prePNe are ecplained in sectdion 5. Finally, we draw general conclusions about the character of the flows that have shaped the the lobes.

## 2. Baseline Model

### 2.1 Methodology

The highly versatile code AstroBEAR (Cunningham et al., 2005ApJ...631.1010C, 2009ApJS..182..519C) was used for all of the present simulations. In our case AstroBEAR solves the Eulerian equations of fluid dynamics in a 2-D plane using an adaptive-mesh-refinement ("AMR") method. The kernel of AstroBEAR is a versatile 3-D hydrodynamic solver that can be applied in a wide variety of circumstances, including magnetized/cooling flows and rotating coordinating systems (Huarte-Espinosa et al., 2012ApJ...757...66H). The code does not contain facilities for simulating emission-line images, predicting an emergent spectrum, or following the transfer of emitted radiation through dusty zones.

We performed the computations in one quadrant of a $16 \times 32$ kau plane in which the injected gas is launched along the y-axis into an isotropic and much slower AGB wind (Ivezić & Elitzur, 2010MNRAS. 404.1415I). The grid cells are δ*r* = 500 au in size at time *t* = 0. The wind is presumed steady (i.e., its speed is constant and its density drops as $r^m$, where *r* is radius from the nucleus and $m \approx -2$). For convenience the downstream gas is presumed to be cold and stationary. The grid size, δ*r* = 500 au, was dynamically decreased by the AMR algorithm by factors up to four powers of two (δ*r* = 31 au) in regions where the state variables exhibit steep gradients. (We will justify this choice of grid resolution in sections 2.3 and 2.5.)

---

[1] See the comprehensive review of pPN formation scenarios in Akashi & Soker 2018MNRAS.481.2754A.

*2.2 Design of the Baseline Model*

Our aim is to develop a generic flow, or "baseline" flow model with the geometry and momentum flux needed to create lobes like those of Fig. 1 of the appropriate shape, size, and observed kinematics at 900y[2]. We take the mass and density structure of the AGB wind near the nucleus as known. Over 100 simulations were run before we found a suitable flow model. Table 1 shows the full set of adopted descriptors of the ambient environment and flow used for the AstroBEAR simulations of the baseline model.

Table 1. Parameters and Baseline Model Values

| Parameters[a] | Value |
|---|---|
| Ambient Gas (isotropic "slow AGB winds") | |
| density at nozzle surface $n_{amb,0}$ (cm$^{-3}$) | $10^4$ |
| density radial power-law *m* | –2 |
| launch speed $v_{amb,0}$ (km s$^{-1}$) | static |
| temperature $T_{amb,0}$ (K) | $10^2$ |
| canvas window width (Δx, Δy) | (16, 32) au |
| initial total mass on the canvas (M$_\odot$) | ~0.01 M$_\odot$ |
| Flow at the nozzle (collimated "fast winds") | |
| nozzle radius $r_0$ (A.U.) | $10^3$ au |
| density $n_{flow,0}$ (cm$^{-3}$) | $3 \times 10^3$ |
| injection speed $v_{flow,0}$ (km s$^{-1}$) | 300 |
| temperature at launch $T_{flow,0}$ (K) | $10^2$ |
| Gaussian 1/e taper angle $\phi_{flow,0}$ | 25° |
| injection duration $\Delta t_{flow}$ (y) | 900 |
| mass injection rate $\dot{M}_{flow}$ (M$_\odot$ y$^{-1}$) | $8 \times 10^{-7}$ |
| total injected mass M$_{flow}$ (M$_\odot$) | $7 \times 10^{-4}$ |
| total injected momentum (g cm s$^{-1}$)[b] | $4 \times 10^{37}$ |

[a] Bold: parameters that are specified in AstroBEAR simulations
[b] Values of $10^{37-39}$ g cm s$^{-1}$ are characteristic of fast winds (Bujarrabal et al., 2001A&A...377..868B)

Hereafter we use the term "nozzle" to refer to the interface where the collimated fast outflow first encounters the isotropic ambient environment. The nozzle's surface is presumed spherical with radius $r_0$. Accordingly, entries in Table 1 that describe the fast winds refer to their values on the nozzle at polar angle 0°; that is, $n_{flow,0}$ is the density of the injected gas where the nozzle intersects the *y* axis at a distance $r_0$ from the origin.

A light Gaussian "tapered" conical flow is one in which both the density and the speed of the injected collimated flow along the nozzle surface fall off as Gaussians with polar angle. We adopted this flow geometry used earlier by LS03 (resembling the structure of magnetocentrifugally driven outflows by Shu et al., 1995) because it is a robust way to produce hollow, candle-flame shaped lobes with curved walls of appropriate dimensions. (The Gaussian taper adds only one additional free parameter.) The opening angle of the flow, $\phi_{flow,0}$, is its 1/e width.

The leading edge of the lobe displaces slower ambient gas as it grows. Thus the speed of the lobe tip is only a fraction of the gas that flows through the nozzle. This applies to some extent to all light flows, no matter what the flow geometry.

Untapered (uniform) diverging flows don't work. They produce filled cone-shaped lobes with straight radial edges and a thin spherical leading rim of swept-up gas, or "plug". In 2-D simulations the lobe will resmble a triangular wedge of pizza with circular plug, or "crust" on its outer border. The lobe is not closed and its walls are straight rather than curved. The fast flow never contacts the straight radial side edges of the lobe. Accordingly, its lateral walls are inconspicuous. Moreover, position-Doppler velocity plots, or "P-V diagrams", of the lobes will show no radial speed gradient (see also Appendix A).

The flow in the baseline model is marginally light: at launch the density of the tapered flow on its symmetry axis is one-third that of the ambient gas at that location prior to launch. Obviousy this ratio drops with polar angle. The tapered geometry of the injected momentum flux naturally forms curved lobe walls of displaced ambient gas. The flow-ambient density contrast and the flow speed determine the length of the lobe as it grows.

The choice of the nozzle geometry bears some elucidation. We adopted a round nozzle of radius $r_0$ = 1000 au whose orifice size is determined by the flow opening angle. This radius is large compared to the scale sizes of the stellar mechanisms (e.g., such as close binary systems or the jets from accretion disks, e.g., Chen et al. 2018MNRAS. 473..747C, Cheng et al 2018ApJ...866...64C) that might produce the fast collimated flow. So turbulence and flow asymmetries within the flow are not included in our models.

This value of $r_0$ is simply an expedient. It was used to assure that the interface between the gas exiting the nozzle and the AGB wind downstream is well resolved with the coarse grids (δ*r* = 31 au) that we adopted to make this study feasible. This is a benign size scale for studies of images of prePNe with the best available spatial resolution, ≈0.″1.

*2.3 Mass and Momentum*

The mass injection rate of the fast collimated wind though the nozzle, $\dot{M}_{flow}$, is determined by its density, $n_{flow,0}$, its speed, $v_{flow,0}$, the nozzle area $4\pi r_0^2$, and the flow opening angle $\phi_{flow,0}$). For the baseline flow (Table 1) the mass injection rate $\dot{M}_{flow} = 8 \times 10^{-7}$ M$_\odot$ y$^{-1}$, or ~7 × 10$^{-4}$ M$_\odot$ after its duration, $\Delta t_{flow}$, of 900y. This agrees with empirical data for AGB stars (Figs. 6, 7, 17, & 18 of Höfner and Olofsson 2018A&ARv..26....1H).

Empirical estimates of the present mass of fast flows are much larger than this, ≈0.01 to 1 M$_\odot$, as measured by

---

[2] We choose a final time of 900y for convenience. At later times the flow exceeds the computational canvas, 16 × 32 kau.

$^{12}$CO fluxes of broad line wings by Bujarrabal et al. (2001A&A...377..868B). This disparity is readily resolved. The $^{12}$CO estimates of fast-wind mass are based on the presumption that all of the gas represented in the high-speed wings of $^{12}$CO lines is freshly injected. In contrast, the baseline model shows that the bulk of the mass in the walls of the lobes is displaced ambient mass that steadily acquired momentum at the lobe walls from the fast tapered flow. Note that our estimated flow momentum, $4\times10^{37}$gm cm$^{-2}$ s$^{-1}$, is typical of the momenta observed in the $^{12}$CO profile wings by Bujarrabal et al.

The total mass of the slow ambient AGB "wind" is 0.014 M$_\odot$ within a radius of 32 kau. The mass of $^{12}$CO of the AGB wind of a prePN at a distance of 2 kpc that lies within a 1′ beam of a radio telescope is about four times as large. The total mass of the AGB wind can be much larger, depending on its outer radius $R_{wind}$ (≈0.4 M$_\odot$ for $R_{wind}$ = 1000 kau). Höfner and Olofsson estimate lifetime AGB wind masses ≳0.25 M$_\odot$ for ≳1 M$_\odot$ stars.

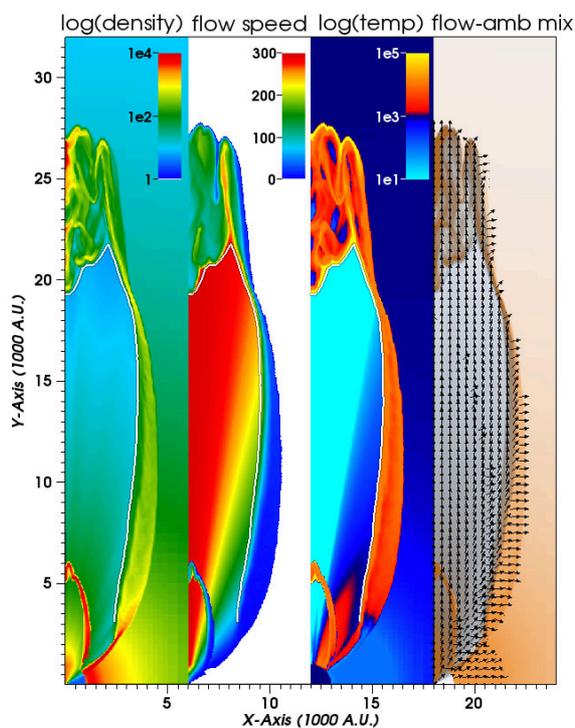

Fig. 3. The "baseline" hydro model based on the parameters in Table 1 and used as a standard of comparison for other simulations throughout this study. The panels show the density, speed, temperature, flow vectors, and the mix of injected (grey-blue) and ambient gas (orange) at 200 and 900y. White lines show the locus of the inner (aka reverse) shock at the terminus of the streamlines.

### 2.4 Salient Features

As shown in Fig. 3, the outcomes of the basline simulation after time $t$ = 900y include a pair of hollow lobes that form on the symmetry axis with an aspect ratio of about 4:1 (Fig 1). The lateral walls are mildly curved. The tips have proper-motion speeds of ≲140 km s$^{-1}$. The flow unit vectors lie along the inside edges of the lobes whereas they are largely horizontal on the outside edges.

Other highlights of the simulation include the attributes and features listed below.

1. The radial streamlines of the flow stream freely through the lobe interior until they encounter and shock the displaced gas at the nebular perimeter (the locus of this "inner" or "reverse" shock is shown as a white line in the first three panels of Fig 2.).
2. The slow ambient gas at the head of the lobe almost immediately slows the leading edge of the lobe to less than 50% of the flow speed at the nozzle.
3. Leading and reverse shocks are separated by a slowly advancing contact discontinuity ("CD").
4. Gas upstream (downstream) side of the CD consists almost exclusively of gas from the injected fast flow (ambient AGB gas) with relatively little mixing.
5. The freely streaming fast gas reaching the front (top) of the lobe overtakes and shocks the thin rim of displaced gas, forming prominent "thin-shell" instabilities[3] at high latitudes. This creates "soft spots" in the walls at high latitude into which the hot post-shock gas penetrates.
6. The instabilities wrinkle or corrugate the leading edges of the lobes, eventually forming one or multiple growing "fingers", each of them driven by the thermal pressure of post-shock gas. (A multi-polar or "starfish" shape can develop; see section 3).
7. Axial knots or "spears" of dense gas develop near the *y* axis (see below for a discussion of their formation).
8. As found by LS03, streamlines entering the lobe walls at oblique angles retain their transverse speed while their forward speeds rapidly slow.
9. As a direct result, streamlines of the post-shock gas follow lobe walls, especially at high latitdues.
10. The lateral edges of the outer portions of the lobe walls (beyond the CD) are driven into the ambient gas by thermal pressure (expansion) of gas. The heat between the CD and the outer walls was originally generated at the tip of the flow as it progressed into the ambient gas and fell behind. This gas thermally expands in place as the tip of the lobe moves onward.
11. Although the streamlines of the injected flow are radial, the flow direction in the zone between the reverse (leading) shock and the CD is largely mesial (lateral).
12. Optical emission lines or, in extreme cases, soft x-rays can appear in post-shock gas where the emisison lines can be thermally excited.
13. The axial ratio of the lobes changes little after 100 y. This outcome is not an artifact of the adopted grid resolution (section 2.5).

To some degree these attributes and features characterize every other model presented in this paper as well as the case studies presented earlier in this series of papers.

---

[3] Thin-shell instabilities are even more important in analogous models without a taper imposed on the injected flow and in 3-D models.

The diverging streamlines in the tapered flow are deflected as they enter and shock the lobe walls (items 9 and 10 above). As first discussed by Frank et al., (1996, ApJ, 471, L53), Canto et al., (1988A&A...192..287C), Tenorio-Tagle et al. (1988A&A...202..256T) and in more detail in LS03, the faat gas will re-stream along the lobe walls and converge to the symmetry axis where it encounters the mirror-image flow coming from the adjacent quadrant. A cooling knot or spear-like structure forms very close to the *y* axis which often becomes denser or changes in shape as the lobe moves forward. This process is especially important at first when the flow is most influenced by its denser surroundings.

The axial knots and spears are robust outcomes of every model in this paper, previous simulations in this series, Dennis et al (2008ApJ...679.1327D), and LS03, whenever the lobes' leading edges are at least partially resolved. However, they may be largely artifacts. They may not form if the fast winds are turblent or irregular. In addition they may be unstable against perturbations in 3-D simulations. Their structure depends on grid resolution (section 2.5).

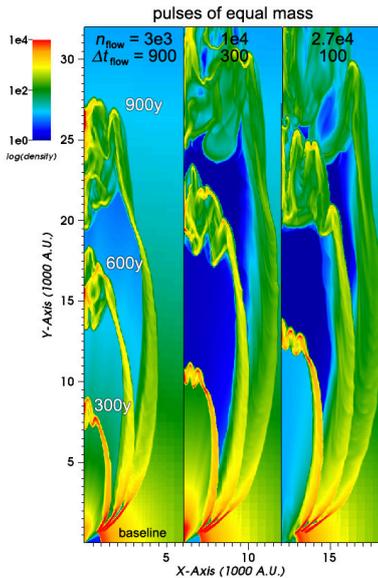

Fig. 4. Alternative baseline models in which a steady flow (left panel) is replaced by 300-y (middle) and 100-y (right) pulses of injected gas, each with the same injection geometry and momentum.

The axial knots and filaments have very few counterparts in actual images of prePNe. The axial knots N3/S3 in M2–9 are the best cases (Balick et al. 2018). Faint axial knots deep inside M1–92 (Fig. 1) may another example.

The paradigm of a light, steady, tapered flow isn't necessarily the only one that serves to create hollow, closed lobes with curved lateral walls. As pointed out by Akashi & Soker (2008 MNRAS.391.1063A), pulsed injections can also form hollow closed lobes. Accordingly, we reran the basline model with pulsed winds of duration $\Delta t$ and equal injected masses and momentum fluxes. The density of the injected gas and the duration of injection were $(n_{flow,0}, \Delta t) = (1\times10^4$ cm$^{-3}$, 300y) and $(2.7\times10^4$ cm$^{-3}$, 100y). A comparison of lobes formed by steady and pulsed winds are shown in Fig. 4. The outcomes differ primarily by differences in the thicknesses of the lobe walls but not their aspect ratios. The kinematics of the gas within the lobe walls is similar in all cases.

## 2.5 Limitations and Reliability

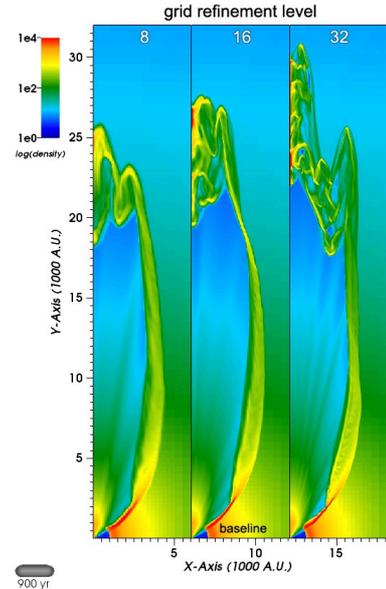

Fig. 5. The baseline model computed at three levels of mesh refinement. These levels correspond to minimum cell sizes of 62, 31, and 16 au.

Constructing numerical models requires a variety of compromises involving limitations of computational speed and model accuracy. We reran the baseline model run with three, four, and five grid refinement levels (cell sizes $\delta r = 500$au/$2^n$, where $n = 3$, 4, and 5; i.e., $\delta r = 63$, 31, and 16 au, respectively). The relative run times are 0.18, 1.0, and 1.8. Nonetheless the large-scale structures of all three models are identical (Fig. 5). The major difference is the complexity of the thin-shell instabilities in the upper quarter of the lobe. Note that $\delta r = 31$ au at 1-2 kpc corresponds to $\sim 0.''02$.

However, running all of the models with the resolution needed to fully characterize the flow emerging from some sort of close binary star system, $\delta r = 1$ au, isn't feasible at this time. Any study at that level of detail requires the incorporation of turbulence, time variations (on the order to the binary period) within the injected fast flow (such as turbulence and variatbility on shorter scales than the binary period).

More relevant here is that the large-scale flow patterns of the baseline model after 900 y are the essentially the same for $\delta r = 63$, 31, or 16 au. So the results presented throughout this paper are generally robust. Even so, the pixelation of the nozzle's surface introduces subtle artificial striations of the propagating flow as it emerges from the nozzle. Such striations can be seen as radial rays in the density structure inside the lobes in Fig. 5. These striations trigger small irregularities at lobe boundaries that slowly grow, depending on the spatial resolution of the simulations.

*2.6 The Formative Years*

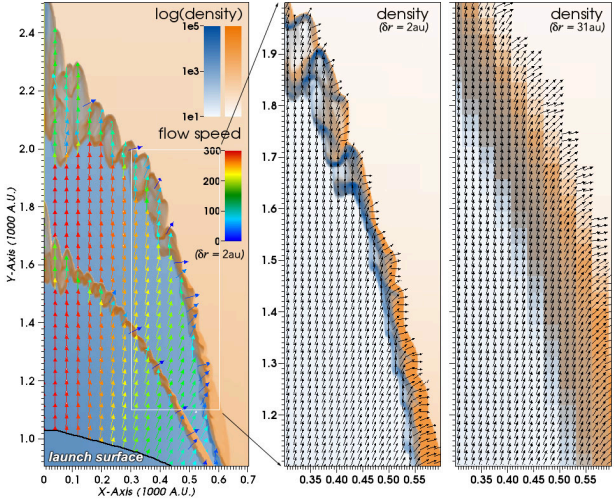

Fig. 6. Details of the flow pattern for the baseline model at two grid resolutions. Left panel: The logarithmic density and velocity patterns of gas at $t = 50$y derived using the baseline model parameters but $\delta r = 2$. The distributions of densities that originate in the injected (ambient AGB winds) are seen in grey-blue (orange). Right: A magnified portion of the lobe walls for $\delta r = 2$ and 31 au. Black unit vectors indicate directions of gas motions. Densities are displayed linearly.

The most rapid transition from initial post-launch to final lobe geometry occurs in the first 50–100 y when the light wind is most effectively influenced and deflected by the ram pressure of the dense ambient gas at the nozzle edge. Detaikls of the rapid changes in lobe structure and flow patterns are illustrated in the two leftmost panels of Fig. 6 at $t = 25$ and 50 y. These frames are computed at a spatial resolution $\delta r = 2$ au (0.2% of the launch radius). Shear instabilities rapidly form and grow along the lobe walls. These instabilities incite rapid and highly localized fluctuations in flow direction, so the edges of the lobe become turbulent. After 50 y the instabilities slide upward and diminish in strength while maintaining their relative spacings as the lobe grows (not shown).

For comparison, the rightmost panel shows the baseline model at $t = 50$y with $\delta r = 31$ au. Comparing the center and right frames shows that the large-scale structure and kinematics of the lobes are unaffected by the instabilities and turbulence along the lobe's lateral edges. We see that the lateral walls slowly but steadily push outward at about 10 km s$^{-1}$. At the same time the forward speed of the lobe tips at 50 y, ~120 km s$^{-1}$, barely changes as the lobes grow. Thus the aspect ratio of the lobe has alrgely settled into its final value.

The main point is that the large-scale shape of the lobe is predictably established by $t \approx 50$y even with the coarse grid used for the lobe shape studies in section 2.3. Only the thicknesses of the lateral walls is poorly represented in models calculated with $\delta r = 31$ au. With this exception, the results presented in the next section—also calculated with $\delta r = 31$ au—are reliable.

## 3. Parameter Variations

In this section we present how the outcomes of the baseline model are affected by variations of the parameters that describe the ambient gas and fast winds. The properties of the fast flow and its environment are described using the values in section 2 (Table 1) unless noted otherwise. All models are run for 900y using a resolution of $\delta r = 63$ au, though we adopt $\delta r = 31$ au for the first 200 y.

*3.1 Variations in the Properties of the Ambient Gas.*

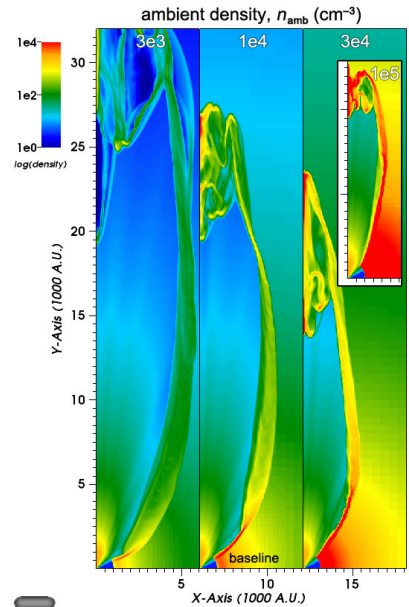

Fig. 7. Models for ambient densities of $3\times10^3$, $10^4$ (baseline) $3\times10^4$, and $10^5$ cm$^{-3}$ at the inner edge of the nozzle, $r_0 = 10^3$au.

*3.1.1. Variations in $n_{amb,0}$.* We first investigate outcomes as the properties of the ambient gas are varied. As expected and seen in Fig. 7, changes in the density $n_{amb,0}$ at $r = r_0$ have major consequences for lobe dimensions, average growth rates, and lateral wall thicknesses. The lobe lengths of the four models are in the ratios of approx. 4/3, 1, ¾, and ½, respectively, while the lobe aspect ratio is only marginally affected. The leftmost lobe may even appear to be open in images of limited signal to noise or poor spatial resolution.

The variations in structure are easily understood since, at fixed momentum flux of the fast wind, the density of the accreted and displaced gas determines the rate of lobe growth. Any changes in shape will be subtle ones, mostly along the leading edge where the amplitude of thin-shell instabilities is sensitive to the ambient density upstream.

*3.1.2. Variations in Density Power Law m.* Changes in the radial power law $m$ of the ambient density distribution $n_{amb}(r)$ from the standard baseline value $m = 2$ have mimic changes in $n_{amb,0}$ (cf. Figs 7 and. 8). (Of course, the initial mass on the grid is affected by changes in *m*. Moreover, departures of *m* from 2 inply that the mass flux of the AGB wind was not historically steady.)

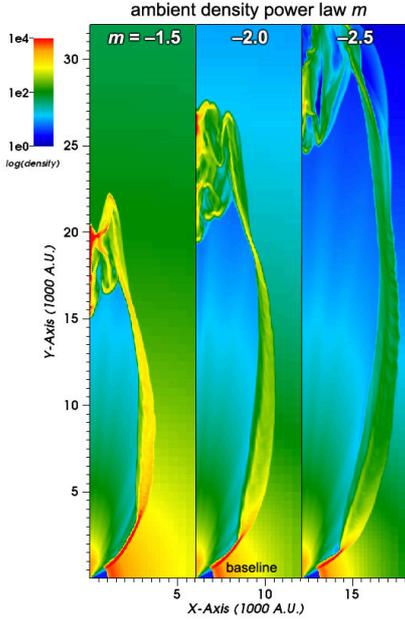

Fig. 8. Models for radial power laws of –1.5, –2 (baseline) and –2.5 in the ambient density distributions, respectively.

*3.1.3. Variations in the Radial Speed and Sound Speed of the Ambient gas.* For computational convenince we assumed that the AGB wind is static. However, we ran baseline models in which its flow speed, $v_{AGB}$, is 10 and 20 km s$^{-1}$, as indicated by dozens of CO spectral observations. This has no appreciable effect on the outcome (not shown). Similarly, the outcomes are very similar so long as the sound speed of the ambient gas is far less than that of the collimated gas. This pertains for $T_{amb} \lesssim 10^5$K.

*3.2 Variations in the Properties of the Injected Gas*

Next we investigate how changes in key flow parameters, density $n_{flow,0}$, injection speed $v_{flow,0}$, momentum flux ($n_{flow,0}\,v_{flow,0}$), and opening angle $\phi_{flow,0}$, affect the model outcomes.

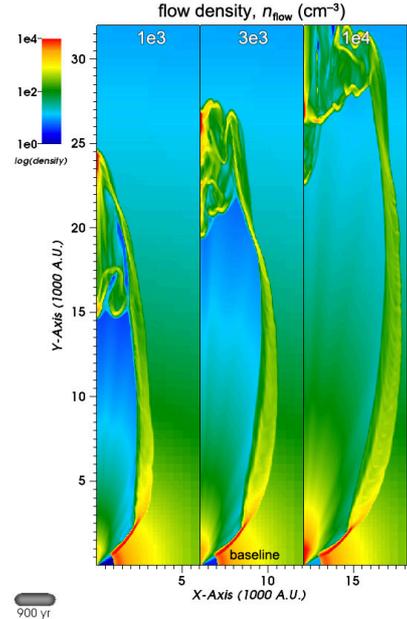

Fig. 9. Models for collimated flow densities of $10^3$, $3\times10^3$ (baseline) and $10^4$ cm$^{-3}$ at $r_0$.

*3.2.1. Variations in $n_{flow,0}$.* All else being equal, denser flows carry more momentum and serve as more effective pistons in displacing the ambient gas (Fig. 9). IRAS 17150-3224 (Fig. 1) is an example of a relatively dense flow. The overall dimensions, shapes, and wall thicknesses of the lobes if are strikingly similar to those in Figs. 7 and 8. All of them are difficult to distinguish observationally.

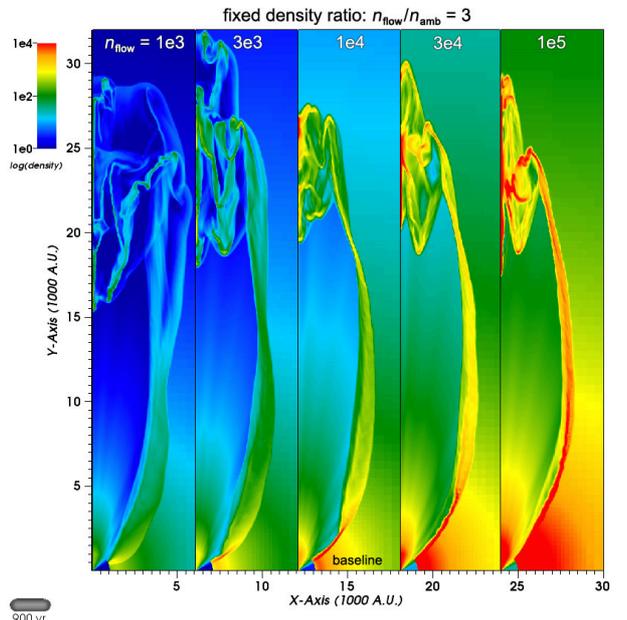

Fig. 10. Models for which the ratio of flow to ambient densities are fixed yet the both densities are varied by factors up to ten. Note how the wall curvature changes and how the lobes morph from open to closed.

*3.2.2. Density Variations that Preserve the $n_{amb,0} : n_{flow,0}$ Ratio.* A comparison of Figs. 7 and 9 suggest that the ratio of flow to ambient densities defines a family of very similar outcomes for a fixed flow speed. This is confirmed in Fig. 10 where the densities are varied by factors of up to ten but their ratios are fixed. The lobes are increasingly closed and instability-induced irregular structure are less important as $n_{amb,0}$ rises.

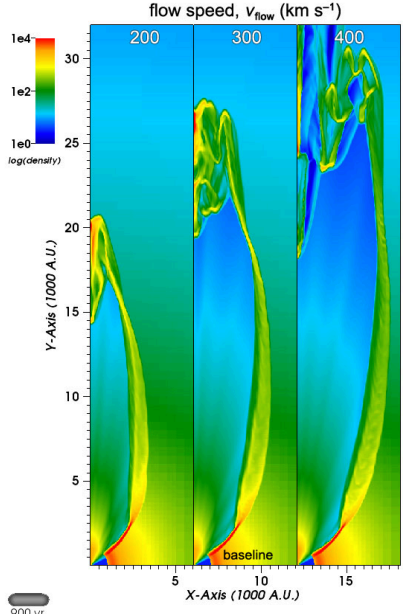

Fig. 11. Models for collimated flow speeds of 200, 300 (baseline) and 400 km s$^{-1}$ at $r_0$.

*3.2.3. Variations in $v_{flow,0}$.* As expected, faster flows pene-trate further into and displace more of the ambient gas (Fig. 11). The lobes formed by faster flows are increasingly longer and wider, much like the outcomes in Fig. 9. Note that the lobe length is not proportional to the initial speed of the flow, as one might naively expect, because the wider lobes formed by faster inflows displace more AGB gas. Note also that thin-shell instabilities are strongly enhanced by higher flow speeds.

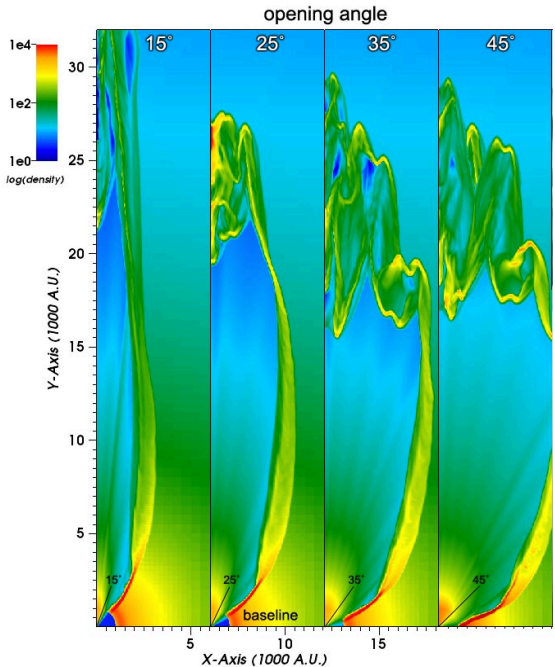

Fig. 12. Models with nozzle opening angles of 15˚, 25˚ (baseline), 35˚, and 45˚.

*3.2.4. Variations in Flow Opening Angle.* Finally we present a series of easily distinguishable models with opening angles of 15˚, 25˚ (the baseline), 35˚, and 45˚ (Fig. 12). Each fast flow has the same momentum flux per steradian at the nozzle launch radius $r_0$. Thus each injected flow makes lobes of about the same overall length. However, the total injected flow mass and momentum scale with the surface area of the orifice.

The tapered flows with small opening angles ($\phi_{flow,0} \leq 15˚$) form lobes with sharp tips that displace ambient gas sideways (like a rocket nose cone) and penetrate fastest into the ambient gas. In direct contrast, the leading edges of light flows with larger opening angles universally exhibit a thin leading rim, or "plug", of swept-up ambient gas whose forward speed is a fraction of that of the flow launch speed (~40% for the light-flow baseline case).

The leading plug is very susceptible to the formation of thin-shell instabilities that will cause it to wrinkle and corrugate (Fig. 12). When the lobe opening angle is large the wrinkles eventually fragment into fingers. This forming "multi-polar" prePNe in which each finger is driven forward by the ram pressure of the streamlines or, more typically, thermal pressure of shock-heated gas from below. The ultimate outcome is a "starfish" morpholgy (Sahai & Trauger 1998AJ....116.1357S). It is useful to compare the respective leading edges of the narrow opening-angle flows of IRAS19255+2123 and IRAS22036+5306 with wide-angle flows IRAS1750-3224 and IRAS19024+0044 in Fig. 1.

## 4. Magnetic Shaping

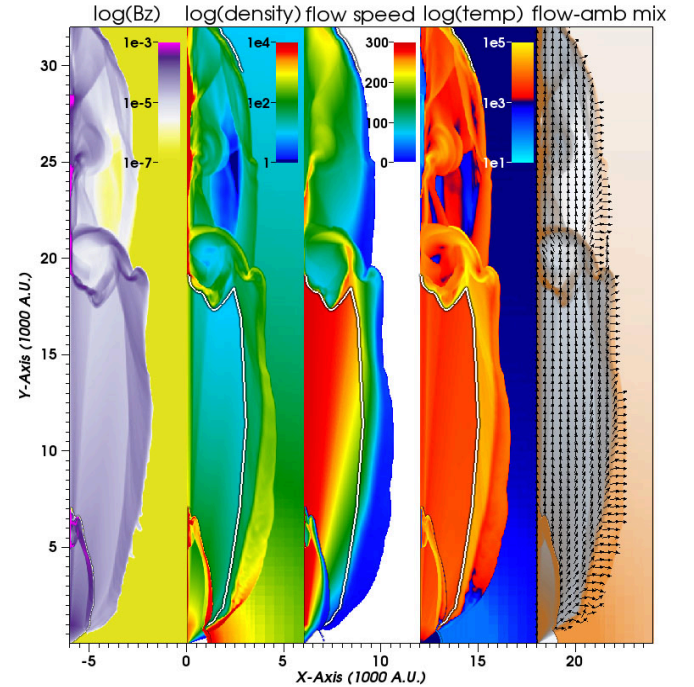

Fig. 13. Similar to Fig. 3 with the exception that the 600-y frames are omitted (for clarity) and a modest toroidal magnetic field of 10$^{-3.5}$ G is added to the flow at the nozzle (far left panel).

## 4.1 Baseline Models with Toroidal Fields

We reran baseline models after injecting toroidal fields of $10^{-5}$, $10^{-4}$, $10^{-3.5}$, $10^{-3}$, and $10^{-2}$ G. Aside from the injected field, the flow is the same as the baseline model. The weakest of fields makes no discernable change in the final lobe morphology from the baseline model in section 2. The strongest value of the field, $10^{-2}$ G, channels the entire injected flow into a very narrow axial cylinder, or "thin jet", containing all of the injected gas. This thin jet is not well resolved in the models.

The most interesting outcome is the magnetized baseline model in which the field is moderate, $10^{-3.5}$ G (Fig. 13). The field channels the flow at $y \geq 20$ kau into the dense, thin, lumpy, and cold jet within which the speeds vary from 300 km s$^{-1}$ at the base to about 200 km s$^{-1}$ at the uppermost tip. The jet appears to pull away from the more slowly expanding lobe over time. In addition the magnetic field carried by the streamlines collects in the walls between the reverse shock (again shown as a thin white line) and the CD where thermal pressure thickens the lobe walls relative to the baseline model.

A white line in Fig. 13 again shows the locus of the inner shock where the radial streamlines terminate. The Doppler shifts from emission lines formed in this post-shock region along the lateral walls show a steadily increasing velocity-distance relationship resulting in a tilted P-V diagram, much as the field-free case (and for the same reasons).

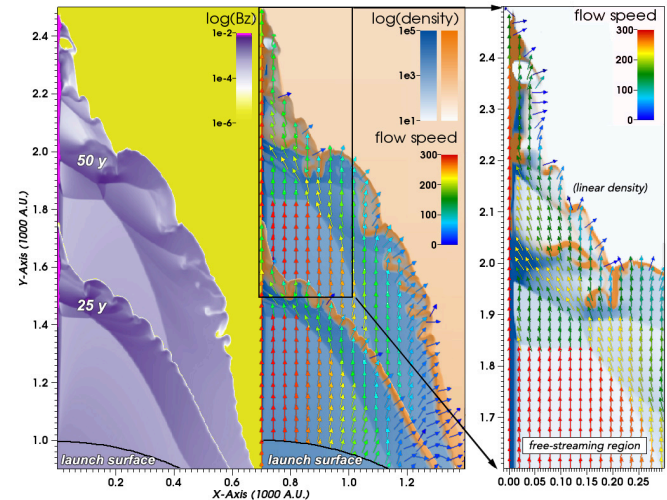

Fig. 14. The outcomes of the high-resolution ($\delta r = 2$ au) toroidally magnetized model at $t = 25$ and 50y. The distribution of the field strength is shown in the left panel. The corresponding density distribution and velocity field are shown in the central panel. As in Fig, 5, gas entering the grid at the nozzle is shown in blue and ambient gas in shades of orange. Speed and flow directions are shown with color-coded unit vectors. The right panel shows an enlargement of the upper portion of the flow at 50y in which density is plotted linearly.

Fig. 14 presents the field strength, mass density, and flow speed patterns at 25 and 50 y at very high spatial resolution ($\delta r = 2$ au). This figure is analogous to field-free baseline model of Fig. 6 in section 2.5. The transisiton from lobe to jet occurs in a thickened leading zone about 20 kau from the nozzle where the magnetic pressure (a tension) competes with both the rapidly dropping ram pressure of the winds and the shock thermal pressure (both thrusts).

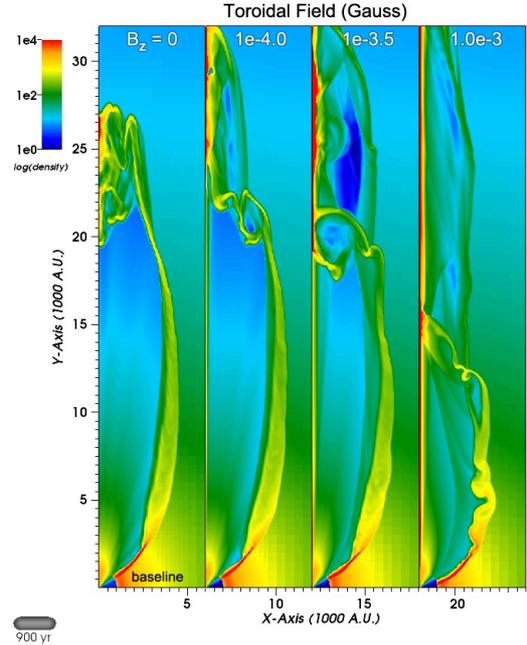

Fig. 15. The density distributions of the baseline model with toroidal fields of 0, $10^{-4}$, $10^{-3.5}$, and $10^{-3}$ G at $t = 900$y.

The strong toroidal field in this zone directs the local gas onto the symmetry axis and thence into the constriction at the base of the thin jet where a strong toroidal field confines it. This flow pattern is best seen in the middle-left region of the center and rightmost panels of Fig. 14 where yellow arrows are superposed on a dark blue background.

Fig. 15 shows the strong relationship between field intensity at the nozzle and overall morphology at 900y. The length and forward speed of the lobe decrease with field strength, yet the prominence and length of the thin axial jet increases dramatically. *However, the inclusion of the magnetic field into the baseline model simply adds complexity to its ultimate structure and significantly decreases the quality of the fit to actual images (Fig. 1).*

## 4.2 Comparison with Previous Models

Pioneering studies of the shaping of hot, ionized PNe by magnetized flows with toroidal fields were published by García-Segura, et al. (1999ApJ...517..767G), Dennis et al., (2009ApJ...707.1485D), and Steffen et al. (2009ApJ...691..696S) of which the latter is most directly comparable to the present models. Steffen et al. considered fast ($10^4$ km s$^{-1}$), uncollimated, nonmagnetized and toroidally magnetized winds (sigma=(0,0.01)) blowing into AGB slow winds of two different forms that were ejected from a rotating AGB star that contains a dense equatorial disk (~$10^5$ cm$^{-3}$ in the midplane).

Other than the injection of a toroidal field into the fast flow (in their case uncollimated), the structure of initial conditions, computational methodology, and spatial resolution adopted by Steffen et al. are all quite different from those used here. Nonetheless, the presence of the toroidal magnetic field in the wind produced basically the same new outcomes: thin, fast, dense "jets" along the symmetry axis behind which trailing wakes form (compare the side-by-side model images in their Fig. 1).

Ciardi, et al. (2013PhRvL.110b5002C), Albertazzi et al. (2014Sci...346..325A), and van Marle, et al., (2014A&A...570A.131V) examined how external poloidal fields (parallel axial field lines) affect the nebulae formed by steady isotropic plasma winds. The growing bubble of gas takes on a very elliptical form with jet-like protrusions along the direction of the field, not unlike our results. NGC7009 also serves to illustrate their model. However, the shaping requires $10^4$ to $10^5$ y before realistic structures form. (van Marle et al.) in typical interstellar fields of ≲5 μG (Haverkorn, 2015ASSL..407..483H ).
 Comfortingly, all of these models end up at about the same endpoint: thin jets that protrude from fairly classical wind-blown structures. They share one key process: the energy density of the winds declines as $r^{-2}$ whereas the magnetic energy density falls off as $r^{-1}$ or slower (depending on the field geometry in the flow). Thus the winds dominate the structure out to a tipping point where the toroidal field takes control of the nebular structure.

*4.3 Possible Examples*

Thin leading, magnetically formed axial jets are uncommon in observations of prePNe. Nonetheless, possible examples of prePNe shaped by magnetized flows include Hen3-1475 (Fang et al., 2018ApJ... 865L..23F), M1-92 (Fig. 1 and Bujarrabel et al., 1998A&A...331.. 361B), and IRAS 19255+2123 (K3-35). Note also that these jets will create a fairly dense wake of displaced ambient gas that resembles the shape of the mature ionized Hen2-428, an ionized PN.

NGC 7009 is an example of a more mature and ionized PN that may have been easily shaped by a toroidal field (Steffen, et al. 2009RMxAA..45..143S) or by a poloidal field (Albertazzi et al.) but probably not by an unmagnetized flow. Other examples include the axial jets in Hen 2–104 and NGC7354, and possibly the ansae of Fleming 1, Hen 2–90, Hen 2–111, Hubble 4, IC 4593, K4-47, M1-16, M1-66, M2-48, NGC 3918, NGC 6543, NGC6751, and NGC 6881. The long-slit kinematic patterns of the jets can are needed fore verification.

**5. Kinematics**

The most readily availabale data that constrain any models of pPN evolution are their morphologies (from images) and kinemtic patterns (from long-slit or fiber-fed spectrographs). The previous sections of the paper focussed the former, so in this section we discuss kinematic patterns.

*5.1 Common Linear P-V Outflow Pattern*

Imaging spectrographs commonly exhibit linear position-velocity ("P-V") correlations in Doppler shifts observed along a slit placed along the symmetry axis. Slits placed on the lobe symmetry axis show much the same patterns: a trend of linearly rising Doppler shift with offset from the nucleus (that is, a tilted P-V outflow pattern). Good data exist for Hen3-401, IRAS16594-4656, Roberts 22 (Hrivnak et al., 2008ApJ...688..327H), M1–92 (Alcolea et al, 2008Ap&SS.313..235A), M2–56 (Castro-Carrizo et al., 2002A&A...386..633C, Sánchez Contreras et al., 2010ApJ...715..143S), and OH231.8+04.2 (Alcolea et al., 2001A&A...373..932A, Sánchez Contreras et al., 2015A&A...577A..52S, 2018A&A...618A.164S. There is often evidence of line splitting attributable to the lateral expansion of the lobes at speeds of about ±10–15 km s$^{-1}$.

Possibly the best spectroscopically observed of all bipolar pPN-like objects is M2–9. M2–9 is unique since long-slit spectra show two distinct patterns: the usual split linear P-V pattern in optical lines and [FeII] λ1.64μm from the inner of two nested lobes plus a pattern of motions with equal, opposite, and constant speed in H$_2$ λ2.12μm from the outer lobes (Smith et al., 2005AJ....130..853S). Balick et al 2018 incorporated these disparate kinematics into one model that is similar in character to the baseline model of section 2.

*5.2 Explaining the Pattern*

Until recently the tilted P-V pattern seen long the walls of the lobes of perPNe was interpreted in terms of ballistic outflows, as if the lobe was mystically created in a brief, energetic event and flung outward without any further change in ishape (cf., Corradi et al., 2001ApJ...553..211C, Alcolea et al., 2001A&A...373..932A, Alcolea et al., 2008Ap&SS.313..235A, Akashi & Soker 2008MNRAS.391. 1063A). This paradigm nicely accounts for the tilt of the P-V diagram but no other attributes. Here we argue that steady (or pulsed) light flows from a central collimator naturally account for the generally closed boundaries of lobes, their curved edges, their hollow interiors, and their tilted P-V patterns in a single paradigm.

We summarize a simple and robust explanation for the common occurrence of tilted P-V diagrams seen along the outer walls of prePNe (refer to Balick et al. 2018 for a much more detailed disussion). The basic ideas are presented in sections 2 and 3 and they apply to all of our simulations whether or not magnetic fields are not present.

White lines in Figs. 3 and 13 show the locus of the inner (reverse) shock where the flow streamlines strike and shock the compressed rim of displaced gas along the lobe walls. This shock is where the flow vectors shed a portion of their kinetic energy and abruptly change direction from purely radial to trajectories that follow—or nearly follow—the rim itself. This is best seen in the orientations of unit vectors located between the reverse shock and the CD in the central panel of Fig. 3. We refer to the post-shock recombination zone where emission lines are collisionally excited (and some species may be collisionally ionized) in the heated post-shock flow.

There are two complementary reasons for the speed gradient in the post-shock zone. Most importantly, the speeds of the streamlines of the tapered flow decrease as a function of polar angle. Thus a plot of streamline speeds immediately inside the inner shock along the sidewalls of the lobe will show a strong and rising gradient. Of comparable significance, the angles at which the streamlines strike the lobe walls are increasingly oblique at higher latitudes along the lobe walls (no matter whether the flow is tapered). These kinematic patterns are a robust and ubiquitous feature in all models.

As noted by Canto et al (1988), Frank et al (1996), and LS03, most of the forward momentum of streamlines entering a shock is dissipated in heating the shock. However, the transverse component of momentum is unchanged. So the flow streamlines veer in angle towards the local walls. (They may not precisely follow the geometry of the walls, but this isn't consequential.)

The tilt of the P-V diagram pattern changes as the lobe grows such that its general slope (length/proper-motion speed) is always a good measure of its age (like $1/H_0$, where $H_0$ is the cosmic Hubble slope.) The fixed geometry of the tapered flow —and the resulting uniformly expanding lobe shape after a few hundred years—assured this outcome. The results are much the same when the fast wind is pulsed rather than steady (section 2.5 and Fig. 4).

(We note that the same tilted P-V trend is often seen in fully ionized PNe. Our explanation obviously doesn't pertain to PNe with simple round or elliptical geometries. On the other hand, at least some bipolar PNe also show P-V diagrams with an overall tilt that is probably vestigial from the time when their lobes formed. Hen2-104 (Corradi et al. 2001ApJ...553..211C) and NGC6302 (Santander-García et al., 2015A&A...573A..56S) are probably the most compelling cases.

### 5.3 Kinematics of Magentic Models

Magnetic models result in a normal but stunted lobe formed by fast, collimated winds plus a thin, jet-like protrusion where the fields dominate the local energy density. The internal characteristics of the lobes formed by light tapered flows are the essentially same as the unmagnetized model. Long-slit observations of the lobe edges should show the normal linear P-V pattern.

However, the kinematics of the dense flow within the thin axial jet upstream of the lobe will not have a tilted P-V diagram in our simulations. The strong surrounding toroidal field squeezes and guides that gas entering the base of the jet to higher speeds (up to $\approx 300$ km s$^{-1}$) before the flow slows from the accumulation of ambient gas near its outer tip (Fig. 14).

## 6. Conclusions

### 6.1 The Lobes of Light Field-Free Outflows

The candle-shaped, hollow lobes of prePNe have lengths $\gtrsim 20$ kau (0.1 pc) and ages $\approx 10^3$y, so the typical speed of the lobe tip is $\approx 100$ km s$^{-1}$.

This paper is one in a series of hydrodynamic studies of the evolution of the lobes of prePNe intended to probe their histories, as constrained by high-quality images and high-dispersion spectra of their lobes. All of the successful simulations in our studies consist of lobes formed by marginally light and tapered flows of various geometries and ejection speeds that enter the spherical environment of previous AGB winds. In this paper we first developed a "baseline" model that fits the attributes of common prePNe. The values of the resulting model parameters are shown in Table 1. We discussed the major features of the baseline model (Figs. 3–5) and how the outcomes may change for pulsed (non-steady) flows (Fig. 6). Then we varied the parameter values in Table 1 to see how the lobe geometry and kinematics are affected (Figs. 7–12) with other parameters kept fixed. Finally we added a range of toroidal magnetic fields to the baseline flow and found that in every case the fit of the models significantly degraded with increasing field strength.

As expected, the lobes of the baseline model shown in this paper (Figs. 3 to 12) show a narrow range of lengths and shapes for the paradigm that we adopted. Lobe widths are primarily functions of the flow opening angle and speed. The lengths are determined by the momenta of the injected flow and the density of the environment. The lengths of lobes are differentiated largely by flow speed, the ratio of flow to ambient density at the surface of the nozzle, and the nozzle opening angle. In addition, kinematic ages determined from lobe length to growth-rate ratio, $\theta/\dot{\theta}$, are reliable unless the lobes projection angle is extreme.

### 6.2 Generalizations of Findings

We exploit this opportunity to compile and summarize the relevant findings about the hydrodynamics of lobe evolution from all of the papers in this series.

1. Hollow, closed lobes are formed by light flows ($n_{\text{flow},0}/n_{\text{amb},0} \approx 0.1$ to $0.3$) that are supersonic relative to the cold and much slower AGB winds into which they are injected.
2. The low specific-momentum gas that they displace quickly retards the growth rate of the lobe tip by 40–60% for such light flows. The retardation scales with the density contrast $n_{\text{flow},0}/n_{\text{amb},0}$.
3. The leading edge of a light flow effectively becomes heavy once it reaches the outer, lower-density regions of the AGB winds. In the baseline model this occurs within about the first 100-200 y.
4. The lobes retain their large-scale shapes (i.e., aspect ratios) once this occurs.

5. The lobe tip can slowly speed up as the density of the upstream gas declines; however, it never regains its initial speed.
6. Flows of limited duration but the same injection geometry and total momentum as a steady flow produce lobes that are very similar in appearance and most likely not distinguishable observationally.
7. The radial streamlines of the injected fast wind flow freely until they reach the walls of swept up and displaced ambient gas at the lobe perimeters.
8. The streamlines of tapered flow are deflected at the compressed, shocked gas in the inner walls of the lobes where they produce slower flows that follow the curved lobe walls (LS03). This produces curved lobe walls in which the flows converge to form closed, flame-like lobes.
9. Shear instabilities along these walls produce local K-H instabilities within which flow and ambient gas do not effectively mix. The vortex-like instabilities spread apart but do not affect the lobe width, length, shape, or speed after they form.
10. Under most conditions thin-shell instabilities quickly arise within the rim of swept up ambient gas at the leading portions of the lobe.
11. These instabilities continue to grow in amplitude once the rims enter the sparse stratified outer regions of the slow AGB winds.
12. The leading edges of wider-angle flows fragment into families of growing fingers that will resemble starfish and multipolar prePNe.
13. The walls of the lobes are bounded by leading (outer) and reverse (inner) shocks separated by a contact discontinuity through which no mixing occurs. However, the CD slowly drifts outward allowing very small amounts of ambient and injected gas to share the same very local volume.
14. The sum of the speeds of the leading and reverse shocks is of order the injection speed of the flow (which is tapered by a Gaussian in our models) in a static ambient medium.
15. Lines of shock-excited species such as [FeII] $\lambda 1.64\mu m$ and low-lying levels of optical forbidden may arise in the recombination zone at the reverse shock. Soft x-rays may arise in extreme cases.
16. The Doppler shifts of these emission lines along the lobe walls will rise steadily with distance. As a result, the P-V diagrams of Doppler shifts from the lobe walls are essentially linear.
17. Gas in the walls outside the CD consists of ambient gas flows that was initially heated to $\sim 10^6$ K by the leading shock. This hot at the lobe edges expands in place as the lobe tip moves beyond it. The expansion is laterally outward (into the ambient medium) owing to the dense gas in the lobe walls.

*6.3 The Lobes of Magnetized Outflows*

We found in section 5 that fast winds containing a toroidal field produces hybrid lobe-jet outcomes (Figs 13–15). When the energy density of the field at the nozzle is modest ($10^{-3.5}$ G for the baseline model), stunted but otherwise ordinary lobes form above the nozzle (where the local energy density is dominated by the ram pressure of the flow from the nozzle). However, the kineitc energy density of the wind falls off as $r^{-2}$ as the streamlines diverge. The field energy density declines as $r^{-1}$ or slower. So at some point the toroidal field takes over, recollimates, and confines the gas at the leading end of the stunted lobe, thereby forming a thin, very dense, and magnetically confined narrow "spear" (with likely internal shocks). The gas flowing into the base of the spear starts out at $\sim 300$ km s$^{-1}$. It then encouters and shocks slower gas downstream, beyond which the confined gas flows slows down to 200 km s$^{-1}$. In section 4.3 we identified several candidate prePNe and ionized PNe in which magnetic shaping appears to have been siginficant.

*6.4 Broader Applicibility*

In 2009 B. Balick, J. Huehnerhoff , and J. Baerny (private communication) compiled and posted images of 74 spatially resolved prePNe from the Mikulski Archives (https://faculty.washington.edu/balick/pPNe/). Of these, 37 have one or more hollow structures with sharp outer edges, 22 have filled protrusions with sharp outer edges, and the rest are difficult to classify. Of those with lobes, 35 are bipolar and six appear to show three or more lobes. Equatorial disks and dark lanes are visible in half of the sample with inclinations that render the disks visible.

Four our purposes, about half of the sample show some evidence for lobes formed by light tapered flows. That is to say, none have radial walls, so their shaping requires an external media with sufficient ram pressures to force the lobe boundaries to be curved.

We conclude that smal generalizations of the flow paradigm of section 2 is the best basis for understanding the evolutionary histories of the formation of prePNe. Indeed, the collimated, tapered light flow concept is the only physics-based process that naturally explains and incorporates the ubiquity of hollow structures in prePNe with sharp and brightened edges and their linear P-V diagrams.

*6.5 Moving Forward*

We have concluded from our simulations that the flows that shape the most common lobes of prePNe are almost certainly simple, light, tapered flows with speeds that are considerably faster than the lobe expansion speeds. However, we have not addressed the origins of these flows since simulations of far greater spatial resolution ($\approx 1$ au) needed to model common envelopes and accretion disks lie far beyond the scope of this paper.

The fundamental question here is a simple one: How does tapering arise? The roots of the tapered flow that we adopted from LS03 go much further back to magnetocentrifugally driven outflows (see section 2.2 and Blackman et al., 2001ApJ...546..288B; Frank & Blackman 2004ApJ...614..737F; Matt et al., 2006ApJ...647L..45M; Huarte-Espinosa et al., 2012ApJ...757...66H). It is an interesting challenge to understand how magnetically formed and impregnated flows are demagnetized as they grow. We require fields of $10^{-5}$ G or weaker on the surface of the nozzle in the baseline model.

The answers await in much earlier times (t < 50 y), intimate size scales (~ 1 au), and using as-yet uncertain and potentially highly variable collimation paradigms (e.g. accretion disks and common envelopes). High-resolution 3-D models open degrees of freedom where instabilities, turbulence, and possible flow asymmetries will thrive and persist (e.g., O'Neill, et al., 2005ApJ...633..717O. This may lead to myriad microscopic ways to tangle and compromise fields. In other words, the answers may be obscured by compexity.

## Acknowledgements

It is a pleasure to thank Jonathan Carroll-Nellenback and the many people in the Physics-Astronomy Computational Group at the University of Rochester who developed and maintain the Astrobear code. We gratefully acknowledge a decade of valuable discussions with Noam Soker and Martin Huarte-Espinosa.

We are thankful support for program AR-14563 that was provided by NASA through a grant from the Space Telescope Science Institute, which is operated by the Association of Universities for Research in Astronomy, Inc., under NASA contract NAS 5-26555. This paper is partially based on observations made with the NASA/ESA Hubble Space Telescope, obtained from the MAST Archive at the Space Telescope Science Institute, which is operated by the Association of Universities for Research in Astronomy, Inc., under NASA contract NAS 5-26555. Support for MAST for non-HST data is provided by the NASA Office of Space Science via grant NAG5-7584 and by other grants and contracts.

**Facility** HST (WFPC2)

**Appendix A**.

The many significant successes of the baseline model include hollow lobes with curved, candle-shaped edges and a common flow patterrn inside the edges of the lateral walls with a linear (or nearly linear) radial gradient of flow speed up to ≈50% of the speed of the injected gas. The curved walls of tapered flows are the direct result of streamlines that obliquely strike the curved lobe walls that they compressed.  Here we argue that uniform, untapered cylindrical or diverging flows from the nozzle fail to account for these highly characteristic shapes and flow patterns.

Untapered cylindrical or diverging flows, whether light or heavy, cannot appear hollow.

In an isotropic external medium untapered flows produce lobes with purely columnar (cylinders) or radial edges (diverging flows) parallel to the outermost streamlines. (Very small shear instrabilities may arise at the flow boundary, as in Fig. 6.)  Thus the lateral walls are not compressed, shocked, or especially conspicuous in scattered starlight. The ambient gas plays an entirely passive role in constraining the walls of the lobes except in the thin, dense rim along the lobes' leading edges.

Obviously there is no mechanism by which ambient gas adjacent to the lateral edges of the lobes can be accelerated to create a P-V diagram that differs in form from that of the AGB wind.

Also, the untapered flows leave behind a wake of previously shocked gas at $10^4$–$10^5$ K from the compressed rim at the leading edge of the lobe. This hot and largely ambient gas spills sideways from the outer corners of the rim and falls behind the lobe as it continbues to grow. There it adiabatically expands in place at its sound speed (≈10 km s$^{-1}$) but has little if any radial motion.

Therefore the P-V diagram from the long-slit spectrum will not produce the tilted pattern indicative of a radial speed gradient (unless there are strong geometric projection effects that enahnce its slope at low inclination).  Instead, the spectrum obtained along a lobe's symmetry axis will resemble that seen for dense ejected clumps such as those found in H-H objects (e.g.. Raga et al., 2004AJ....127.1081R).  The emission lines will flare where the slit crosses the shock-heated rim. Behind the rim the long-slit spectrum will show the ±10 km s$^{-1}$ lateral expansions of the front and rear surfaces of the trainling wake.